\documentclass[usenatbib,galley]{mn2e}
\usepackage{graphicx}

\title{Lynds 1622: a nearby star forming cloud projected on Orion B?}

\author[M. Kun et al.]{M. Kun$^{1}$\thanks{E-mail: kun@konkoly.hu}, Z. Balog$^{2}$\thanks{On leave from University of Szeged,
D\'om t\'er 9, Szeged, H-6720 Hungary},  N. Mizuno$^{3}$, A. Kawamura$^{3}$,  A. G\'asp\'ar$^{2}$, S. J. Kenyon$^{4}$,\\
\newauthor  and  Y. Fukui$^{3}$ \\
$^{1}$Konkoly Observatory, H-1525 Budapest, P.O. Box 67, Hungary \\
$^{2}$Steward Observatory, University of Arizona, 933 N. Cherry Av., Tucson AZ, USA 85721 \\
$^{3}$Department of Astrophysics, Nagoya University, Nagoya, 464-01, Japan \\
$^{4}$Harvard--Smithsonian Center for Astrophysics}
\begin{document} 

\date{Received  / Accepted }

\maketitle

\label{firstpage}

\begin{abstract}  
We present results of optical spectroscopic and photometric observation of the pre-main 
sequence stars associated with the cometary shaped dark cloud Lynds~1622, and 
$^{12}$CO and $^{13}$CO observations of the cloud. We determined the effective 
temperatures and luminosities of 14 pre-main sequence stars associated with 
the cloud from their positions in the Hertzsprung--Russell diagram, as well 
as constructed their spectral energy distributions using optical, 2MASS and
\textit{Spitzer} IRAC and MIPS data. We 
derived physical parameters of L\,1622 from the molecular observations. 
Our results are not compatible with the assumption that L\,1622 
lies on the near side of the Orion--Eridanus loop,
but suggest that L\,1622 is as distant as Orion~B. At a distance of 400~pc 
the mass of the cloud, derived from our $^{12}$CO data, is 1100\,M$_{\sun}$, its 
star formation efficiency is $\sim 1.8\%$, and the average age of its 
low-mass pre-main sequence  star population is about 1~million years.
\end{abstract}

                                                                               
\section{Introduction}
\label{Sect_1}

Studies of the large-scale properties of star forming regions are important 
for understanding the interstellar processes governing star formation.
Census of young stellar objects (YSOs) born in a cloud, their mass, age 
and space distribution, together with the mass of the star forming cloud
are key data for assessment of the efficiency and time scale of star formation. 

The Orion~OB1 association and its associated molecular clouds represent 
a distinguished target for star formation studies, 
being the nearest giant star forming region where various stages of the star 
forming processes across the whole stellar mass spectrum, 
as well as interactions between the consecutive generations of 
stars can be studied in detail. The target of the present paper, 
Lynds~1622 and the neighbouring Lynds~1621 are small dark clouds at the low Galactic 
latitude edge of the giant molecular cloud Orion~B. These two clouds together are termed
as {\em Orion East\/} by \citet{HR72}, who discovered five H$\alpha$ emission stars
associated with L\,1622. \citet{Maddalena} found that the radial velocity  of 
this cloud is about $v_\rmn{LSR} = +1$\,km\,s$^{-1}$, contrary to the 
characteristic radial velocity of +10\,km\,s$^{-1}$ of Orion~B. The asymmetric, cometary shape 
and the bright, ionized rim apparent in the SHASSA \citep{Gaustad} image of L\,1622 
suggest its interaction with the hot stars of Ori~OB\,1. \citet{Casassus} present
centimeter continuum image of L\,1622, and attribute the radiation to spinning dust grains
in a limb-brightened shell of the cloud, where the incident UV radiation from the 
Ori~OB1b association heats and charges the grains.

Low-mass star formation in L\,1622 is indicated by several associated T~Tauri stars 
and candidates. The reflection nebula VDB~62, illuminated by the K2 type weak-line 
T~Tauri star HBC~515 (HD\,288313, V1793~Ori) 
\citep{Racine,HBC} can be found near the bright rim of 
the cloud. Spectroscopic and photometric variability of this star were
studied in detail by \citet{Mekkaden}, and it was found to be a triple system by \citet{Reipurth08}. 
In addition to this WTTS, there are six known classical T~Tauri stars 
(CTTS) in L\,1622: HBC~188 (LkH$\alpha$~334), HBC~189 (LkH$\alpha$~335), 
HBC~190 (LkH$\alpha$~336), HBC~516 (LkH$\alpha$~336c), HBC~191 (LkH$\alpha$~337), 
and HBC~517 (L\,1622--15) \citep{HBC}. The LkH$\alpha$~336 triple system
was studied by \citet{Correia}.  \citet{OH83} detected 16 H$\alpha$ 
emission stars (L\,1622--1 to L\,1622--16) in an objective prism survey over the surface 
of L\,1622, including the above HBC stars except LkH$\alpha$~336c
and HD\,288313. The pre-main sequence (PMS) nature of 11 objects has not yet 
been confirmed.

A further signpost of low-mass star formation is the Herbig--Haro object HH~122 
\citep{Reipurth89}, whose probable exciting source, HH\,122~VLA~1 was detected by \citet{Rodriguez}. 
\citet{LM99} report four optically selected cores of the L\,1621\,/\,L\,1622 region:
L\,1621--1, L\,1621--2, L\,1622~A and L\,1622~B. L\,1622~B is associated with 
the optically invisible source \textit{IRAS}~05522+0146,
while \citet{LMT01} list L\,1622~A as an infall candidate.
Recently \citet{Reipurth_hb} identified 28 young stellar objects (YSOs) in the
\textit{Spitzer} IRAC images of L\,1622, 10 of which they classified  
as protostars.
  
The velocity difference between Orion~B and L\,1622 raises the question
whether both clouds  are located in the same volume of space, or L\,1622 
is a foreground object, unrelated to the Orion~B molecular cloud. 

\citet{Maddalena} concluded that, in spite of the 
different velocities, Orion East is probably located at the same 
distance as Orion~B, regarding its cometary shape and bright rim, 
pointing toward Barnard's Loop. This conclusion was questioned by
\citet{Knude}, who studied the distance to interstellar extinction features by
using spectral types of the Michigan Catalog and Tycho--2 photometry, and 
found an absorbing cloud located 
at some 160~pc from the Sun toward the line of sight of L\,1622, 
suggesting that this cloud may be a foreground object.
\citet{Wilson05}, based on the parallaxes of three \textit{Hipparcos\/} stars,
find that L\,1622 is situated as close as 120\,pc to the Sun. Nearby star forming
clouds are of special interest because they allow us to study fine structural
details and detect the lowest luminosity objects. 

Motivated by these findings we started a spectroscopic and photometric 
study of the pre-main sequence stars and candidates associated with L\,1622.
Our aim is to assess the probable star forming history of the cloud, and see 
if the measured properties of these objects support the smaller or the larger distance. 

In order to have a homogeneous data set on the young low mass stars
born in L\,1622 we obtained moderate resolution optical spectra and $VR_\rmn{C}I_\rmn{C}$ 
photometry of all pre-main sequence stars and candidates projected on L\,1622. 
\textit{Spitzer} archive data were used for studying the spectral energy 
distributions (SEDs) of the young stars.
In order to find the physical properties of the cloud, compare 
the positions of the stars and the molecular gas, as well as estimate the efficiency 
of star formation  we analysed $^{12}$CO and $^{13}$CO observations 
of the cloud available in the data archive of 
the NANTEN radio telescope at Nagoya University. 
Our observations and results are described in Sect.~2, and discussed in Sect.~3.
Section~4 gives a short summary of our findings.

\section{Observations and results}
\label{Sect_2}

\subsection{Spectroscopy}
\label{Sect_2.2}

\subsubsection{FAST spectra of PMS candidates} 

Low-resolution optical 
spectra of the 16 H$\alpha$ emission stars of the field 
of L\,1622, listed by \citet{OH83}, were acquired
with FAST, a high throughput slit spectrograph mounted at the Fred L. Whipple
Observatory 1.5-m telescope on Mount Hopkins, Arizona \citep{Fabricant}. 
In addition to the H$\alpha$ emission stars we observed HD\,288313 too. We observed 
both components of the visual binary L\,1622-10 \citep{RZ93}, 
separated by some 4.4~arcsec.

We used a 300~g~mm$^{-1}$ grating blazed at 4750~\AA, a 3-arcsec slit, and a
thinned Loral 512 $\times$ 2688 CCD. These spectra cover 3800--7500~\AA\ at a
resolution of $\sim$ 6~\AA. We reduced and analysed the spectra in  
{\sevensize IRAF}\footnote{IRAF is
distributed by the National Optical Astronomy Observatories, which
are operated by the Association of Universities for the Research in
Astronomy, Inc., under cooperative agreement with the National Science
Foundation. http://iraf.noao.edu/}.

After trimming the CCD frames at each end of the slit, we corrected for the bias
level and flat-field for each frame, applied an illumination correction,
and derived a full-wavelength solution from calibration lamps acquired
immediately after each exposure. The wavelength solution for each frame
has a probable error of $\pm$0.5--1.0\,\AA. To construct final 1-D spectra,
we extracted object and sky spectra using the optimal extraction algorithm 
within {\sevensize APEXTRACT}. Most spectra have moderate signal-to-noise, $S/N \sim 30$ per
pixel. The spectral classification was performed by visually comparing the 
spectra of MK standards with the spectra of our targets following the 
procedures described in \citet{Jaschek}. We estimate an accuracy $\pm1$ subclass 
for our spectral classification. For the LkH$\alpha$ stars our derived 
spectral types agree with those found by \citet{CK79}. The FAST spectra, normalized 
with the value of the continuum at 6560\,\AA, are shown in Fig.~\ref{Fig_fast}.

\subsubsection{CAFOS spectra of PMS candidates} 

Part of the target stars 
have been observed with the CAFOS instrument on the 2.2-m telescope of Calar 
Alto Observatory in February 2007. Using the grism R-100, the observed part 
of the spectrum covered the wavelength interval 5800--9000\,\AA. The spectral 
resolution of CAFOS observation, using a 1.5-arcsec slit, was 
$\lambda / \Delta \lambda \approx 1000$ at
$\lambda=8500$\,\AA. We reduced and analysed the spectra using standard 
{\sevensize IRAF} routines. 
The wavelength range of   spectra was suitable for determining 
several flux ratios defined as tools for spectral classification of
late K and M type stars by \citet{Kirk} ({\sl A, B, C,
B\,/\,A, B\,/\,C\/}), \citet{MK} ({\sl I$_2$, I$_3$\/}), and  \citet*{PGZ} 
({\sl T1, T2\/}). We measured these spectral features on the spectra of our 
stars, and calibrated them  against the spectral type and luminosity class
by measuring them in a series of standard stars published by \citet{LeBorgne}. 
The accuracy of the spectral classification,
estimated from the range of spectral types obtained from different
flux ratios, is $\pm1$ subclass (for further details of spectral classification
see \citealt{KPNJW}). The spectral types, obtained from the CAFOS observations,
are consistent with those obtained from the FAST spectra.

\begin{figure*}
\centering{\includegraphics[width=16cm]{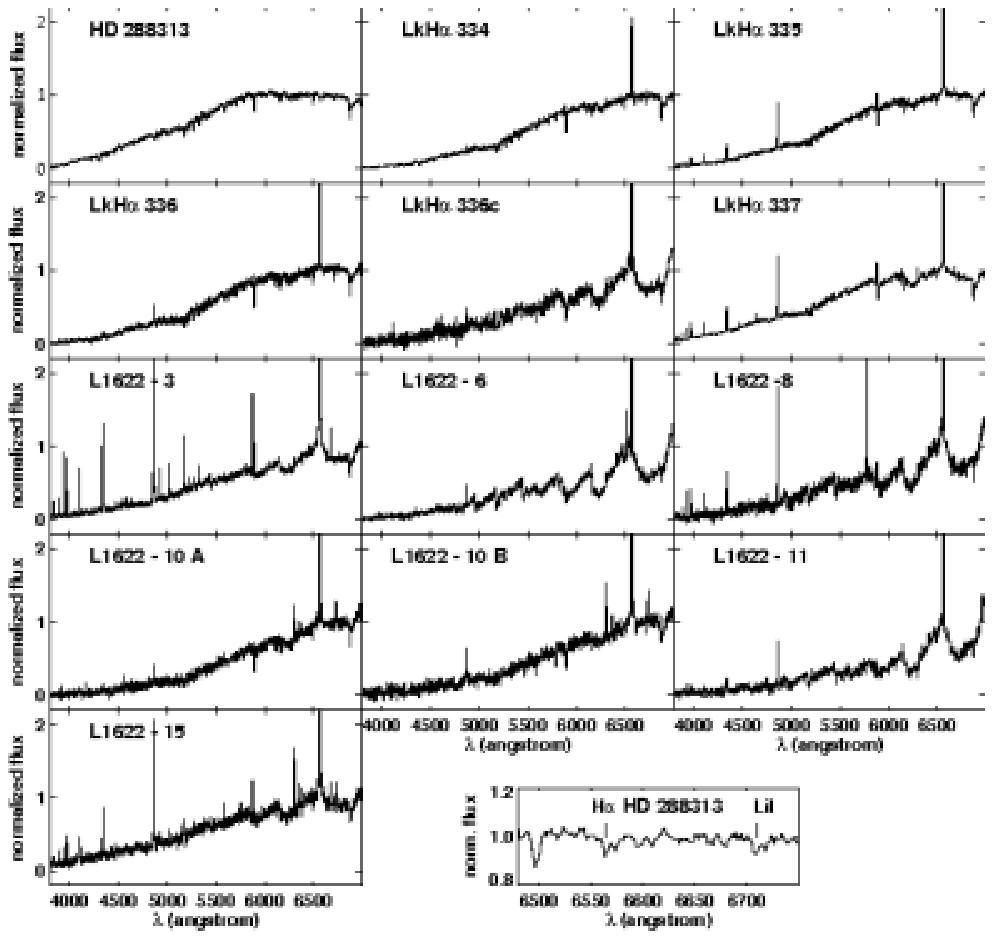}}
\caption{FAST spectra of the pre-main sequence stars associated with L\,1622. The separate 
panel at the bottom shows the part of the spectrum of HD\,288313 around the H$\alpha$ and
\mbox{Li\,{\sevensize I}}.}
\label{Fig_fast}
\end{figure*}

\begin{figure*}
\centering{\includegraphics[width=16cm]{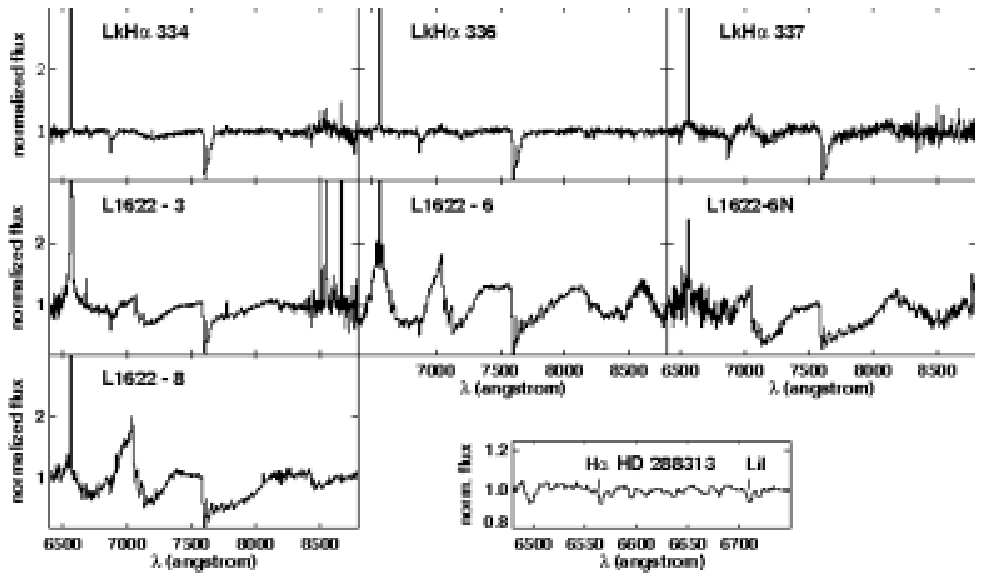}}
\caption{CAFOS spectra of pre-main sequence stars associated with L\,1622. The separate 
panel at the bottom shows the part of the spectrum of HD\,288313 around the H$\alpha$ and
\mbox{Li\,{\sevensize I}}.}
\label{Fig_cafos}
\end{figure*}

We accepted an object as a classical T~Tauri star if it fulfilled the  criterion  
established by \citet{Barrado}, i.e. $EW$(H$\alpha$) exceeded the saturation limit of the
chromospheric activity, depending on the spectral type. The H$\alpha$ emission of 
such stars is thought to originate from the gas accreted by the star from the disc. 
The spectra of L\,1622--1, L\,1622--2, L\,1622--7, L\,1622--13, L\,1622--14, and 
L\,1622--16 have shown no pre-main sequence signature. We noticed,
however, that the spectrum of a faint M-type star, 
2MASS\,J\,05540201+0140558, which appeared in the slit 
next to the CAFOS spectrum of L\,1622--6, has shown W(H$\alpha$) $\approx$ 15\AA.
The critical $EW$(H$\alpha$) for its spectral type M3, according to 
\citeauthor{Barrado}, is about 11\,\AA, suggesting that 2MASS\,J\,05540201+0140558 
may be a CTTS. The equivalent width of the H$\alpha$ emission line of M-type stars,
however, may be slightly overestimated  at the moderate resolution of our observations 
because, due to the TiO band and 6569\,\AA, the position of the continuum is 
somewhat uncertain \citep{WB03}. Taking into account this effect we regarded this star 
as a candidate pre-main sequence object and designated it as L\,1622--6N.   
HD\,288313 is a weak-line T~Tauri star, whose pre-main sequence nature 
is shown by the strong \mbox{Li\,{\sevensize I}} absorption 
\citep[cf.][]{Mekkaden}. Its spectrum shows a weak H$\alpha$ emission,
as the photospheric H$\alpha$ absorption, expected for a K2 type star,
is nearly filled in with emission (see the last panels of Figs.~\ref{Fig_fast}
and  \ref{Fig_cafos}). According to the $EW$(H$\alpha$) values, 12 of 14 PMS 
stars covered by our spectroscopic survey are accreting objects. 

Table~\ref{Tab_sp} shows the final list of the targets which have shown 
spectra characteristic of pre-main sequence stars.  
In addition to the derived spectral types we present the equivalent widths 
of the H$\alpha$ and \mbox{Li\,{\sevensize I}} lines in \AA, 
as well as indicate the additional emission lines observed in the
spectra. The uncertainties given in parentheses have been derived 
from the repeatability of the measurements. We used the {\tt deblend} command of
the {\tt splot} task of {\sevensize IRAF} to separate the \mbox{Li\,{\sevensize I}} line
from the neighbouring absorption or emission features (\mbox{Ca\,{\sevensize I}}\,$\lambda$\,6718, 
\mbox{[S\,{\sevensize II}]}\,$\lambda$\,6717).  The CAFOS spectra are shown in Fig.~\ref{Fig_cafos}.
The strong H$\alpha$ emission lines show that all these objects are classical T~Tauri stars.
L\,1622--3 exhibits an especially rich emission spectrum, characteristic of strongly
accreting or eruptive (EXor type) stars. 
The spectral types of the stars were converted into effective temperatures 
$T_\rmn{eff}$ following \citet{KH95}. 
   
\begin{table*}
\begin{center}
\caption{Results of spectroscopy }
\label{Tab_sp}
{\footnotesize
\begin{tabular}{ll@{\hskip2mm}r@{\hskip1mm}c@{\hskip3mm}l}
\hline 	
Star &  Sp.T. & $EW$(H$\alpha$) & $EW$(\mbox{Li\,{\sevensize I}}) & Other emission lines \\
   &  & (\AA)~~~~  & (\AA) \\
\hline
HD\,288313       & K2    & 0.80\,(0.1)~~     &  0.45\,(0.02)   & $\cdots$  \\
L\,1622--3       & M0  & $-$295.0\,(8.0)~~ &  $\cdots$ & Balmer series, \mbox{Ca\,{\sc ii}} H \& K; \mbox{Ca\,{\sc ii}} IR triplet, \\
 & & & & HeI\,$\lambda \lambda$\,5873, 6678, \mbox{Na\,{\sc i}}, numerous \mbox{Fe\,{\sc ii}} lines \\
LkH$\alpha$\,334 & K4  & $-$8.5\,(0.5)~~  & 0.65\,(0.03) & H$\beta$, H$\gamma$;  \mbox{He\,{\sc i}}\,$\lambda$\,5873, \mbox{Ca\,{\sc ii}} IR triplet  \\   
LkH$\alpha$\,335 & K4  & $-$52.5\,(2.0)~~ & $\cdots$ & Balmer series, \mbox{He\,{\sc i}}\,$\lambda \lambda$\,5873, 6678; \mbox{Ca\,{\sc ii}} IR triplet  \\  
L\,1622--6       &  M4  & $-$20.5\,(1.0)~~ & 0.51\,(0.05) & H$\beta$  \\  
L\,1622--6N  & M3  & $-$14:~~ & $\cdots$ & $\cdots$ \\  
L\,1622-8         &  M2.5  & $-$103.0\,(1.0)~~ & $\cdots$ & Balmer series; \mbox{Ca\,{\sc ii}} H \& K;  \mbox{He\,{\sc i}}\,$\lambda$\,5873   \\  
LkH$\alpha$\,336  &  K7  & $-$26.0\,(1.0)~~ & 0.29\,(0.05) & Balmer series, \mbox{Ca\,{\sc ii}} H \& K \\
LkH$\alpha$\,336c &  M1  & $-$54.8\,(2.0)~~ &  0.53\,(0.05) & H$\beta$, H$\delta$  \\  
L\,1622--10\,A    &  K7  & $-$41.0\,(3.0)~~ & $\cdots$ &  [\mbox{O\,{\sc i}}]\,$\lambda \lambda$\,6300,\,6363; [\mbox{N\,{\sc ii}}]\,$\lambda$\,6583, [\mbox{S\,{\sc ii}}]\,$\lambda \lambda$\,6716,\,6731 \\
L\,1622--10\,B    &  K5  & $-$79.9\,(3.0)~~ & $\cdots$ &  H$\beta$ \\ 
L\,1622--11       &  M2  & $-$65.1\,(1.0)~~ &  0.81\,(0.02)  & Balmer series, \mbox{Ca\,{\sc ii}} H \& K  \\ 
LkH$\alpha$\,337  &  K7  & $-$50.5\,(1.0)~~ &  0.45\,(0.05) & Balmer series, \mbox{Ca\,{\sc ii}} H \& K; \mbox{He\,{\sc i}}\,$\lambda \lambda$\,5873, 6678; [\mbox{O\,{\sc i}}]\,$\lambda$\,6300 \\  
L\,1622-15        &  M0  & $-$72.2\,(2.0)~~ & $\cdots$ &  Balmer series, \mbox{Ca\,{\sc ii}} H \& K;  \mbox{Mg\,{\sc i}}\,$\lambda$\,5172;
[\mbox{O\,{\sc i}}]\,$\lambda \lambda$\,6300,\,6363 \\
 &&&& \mbox{He\,{\sc i}}\,$\lambda \lambda$\,5873, 6678;  [\mbox{S\,{\sc ii}}]\,$\lambda \lambda$\,6716,\,6731; \mbox{Fe\,{\sc ii}} \\    
\hline
\end{tabular}}
\end{center}
\end{table*}

\subsubsection{Spectrum of VDB 63} 

In addition to the candidate PMS stars we observed the spectrum of HD\,288309, the
illuminating star of the reflection nebula VDB~63, located at the eastern edge of L\,1622
(see Fig.~\ref{Fig_map}), on the 4900--7800\,\AA \ region, using CAFOS with the grism G--100. 
The spectral type and photometric data of this star may constrain the distance
of L\,1622. We derived a spectral type K0\,III. The absolute magnitudes of 
K0 giants, tabulated by \citet{Wainscoat}, together with the photometric data 
of HD\,288309 found in the NOMAD\footnote{http://www.nofs.navy.mil/nomad} 
catalogue, result in a most probable distance $D \approx 320$\,pc
for VDB~63. The large dispersion of the absolute magnitudes of K giants, 
$\sigma_\rmn{M}=0.7$ \citep{Wainscoat}, lends a high uncertainty to this value. 
Taking into account the 1-$\sigma$ range of absolute magnitudes, the corresponding
distance interval is $230\,\rmn{pc} \la D \la 440\,\rmn{pc}$.  

\subsection{$\bmath{V R_\rmn{C} I_\rmn{C}}$ photometry}
\label{Sect_2.3}

Photometric observations of the young stars  in the $V$, $R_\rmn{C}$ 
and $I_\rmn{C}$ bands were undertaken on 15 January and 14 February 2007 using 
the 1-m RCC-telescope of Konkoly Observatory. We used a Princeton 
Instruments VersArray:1300B camera, that utilizes a back-illuminated, 
1300$\times$1340 pixel Roper Scientific CCD. 
The pixel size is 20\,$\mu$m, corresponding to 0.31~arcsec on the sky.
Integration times were between 5\,s and 300\,s.
In order to calibrate the observed fluxes, we bracketed the 
observation of each target star by observations of 
four nearby \citet{Landolt} standard stars. We reduced the images in {\sevensize IRAF}. 
After bias subtraction and flatfield correction psf photometry 
was performed using the {\tt phot} and {\tt allstar} tasks of the {\sevensize DAOPHOT} package. 

The results of the photometry  are presented in Table~\ref{Tab_phot}. 
The photometric errors, given in parentheses, are quadratic sums of the formal errors 
of the instrumental magnitudes and those of the coefficients of the transformation 
equations. 

\subsubsection{Interstellar extinction}

The unreddened colour indices of the spectral types were adopted from \citet{KH95}. 
The interstellar extinction suffered by the program stars can be derived from their 
colour excesses. The colour excesses of pre-main sequence stars, however, reflect not 
only the interstellar reddening, but are also influenced by several other effects, such as 
the light, scattered from the disc and envelope surrounding the star, as well as thermal emission 
from the disc and accretion shocks on the stellar surface. It is usually assumed that most of the
$E_{R_\rmn{C}-I_\rmn{C}}$ colour excess results from interstellar reddening \citep[e.g][]{Hartigan,Meyer}, 
and thus it is a reliable estimate of the extinction using the relationship 
$A_V^\rmn{RI}= 4.76 \times E_{R_\rmn{C}-I_\rmn{C}}$ \citep{Cohen}. 

During our data analysis, however, we realized that
the   $E_{R_\rmn{C}-I_\rmn{C}}$ colour excesses of  
LkH$\alpha$\,336c,  L1622--10A, and L1622--10B are probably influenced by 
scattered light: $E_{R_\rmn{C}-I_\rmn{C}}$  apparently underestimates the extinction 
of these stars: the  luminosities, derived from the  $I_\rmn{C}$ magnitudes, corrected for the 
extinction based on $E_{R_\rmn{C}-I_\rmn{C}}$, are too low for classical T~Tauri stars, 
located in the distance interval 120--400~pc from the Sun.
The presence of scattered light is supported by our photometric images,
which show the profiles of these stars slightly broader than the average 
stellar profile of the image. We therefore determined the extinction using the $I_\rmn{C}-J$
colour excess as well, using the 2MASS \citep{2MASS} data, and compared the 
extinctions obtained from both methods.
We found that in most cases both colour excesses result in virtually the same $A_V$, 
suggesting that both the influence of scattered light in the $R_\rmn{C}$ band and
the $J$-band emission from the disc inner rim are negligible.
Exceptions are  the above stars. Probably scattered light makes these stars
bluer in the $R_\rmn{C}-I_\rmn{C}$, and, at the same time, their $E_{I_\rmn{C}-J}$
colour excess is slightly increased by the possible contribution of the disc
emission to the $J$ magnitude \citep{Cieza05}. Therefore we adopt as a best 
estimate of $A_V$ the average of $A_V^\rmn{RI}= 4.76 \times E_{R_\rmn{C}-I_\rmn{C}}$  
and $A_V^\rmn{IJ} = 3.23\times E_{I_\rmn{C}-J}$ for each star. 
The $A_V$ values, as well as the 2MASS designations of the target stars, formed 
from their equatorial coordinates, are also given in Table~\ref{Tab_phot}. 
For determining the extinction in the other photometric bands, we used the
relations $A_{R_\rmn{C}} = 0.78\,A_{V}$ and  $A_{I_\rmn{C}} = 0.59\,A_{V}$  
for the optical  \citep{Cohen}, and $A_{J} = 0.26\,A_{V}$,
$A_{H} = 0.15\,A_{V}$, $A_{K_{s}} = 0.10\,A_{V}$, presented by \citet{NC05} for the 
2MASS-bands. 

Bolometric  luminosities were derived by applying the bolometric corrections $BC_{I_\rmn{C}}$, 
tabulated by \citet{Hartigan} to the dereddened $I_\rmn{C}$ magnitudes, based on the assumption 
that the total emission of the stars in the $I_\rmn{C}$ 
band originates from the photosphere \citep[e.g.][]{Meyer,Cieza05}.
We determined $L_\rmn{bol} / L_{\sun}$ for various distances found in the 
literature.

\begin{table*}
\begin{minipage}{126mm}
\begin{center}
\caption{$V R_{C} I_{C}$ magnitudes and  derived visual extinctions
of the pre-main sequence stars of L\,1622}
\label{Tab_phot}
{\small
\begin{tabular}{lcccccc}
\hline
Star  & 2MASS\,J & $V$ &  $R_{C}$ &  $I_{C}$ & $A_V$  \\
\hline
HD\,288313       &  05540300+0140218 & 10.014\,(0.02) &	  9.263\,(0.02)	&  8.828\,(0.02) & 0.93   \\
L\,1622--3       &  05533298+0217200 & 16.365\,(0.02) &  15.282\,(0.02)	& 14.171\,(0.02) & 1.33   \\ 
LkH$\alpha$\,334 &  05534090+0138140 & 13.122\,(0.02) &  12.365\,(0.02)	& 11.617\,(0.02) & 1.48   \\
LkH$\alpha$\,335 &  05535869+0144094 & 14.201\,(0.02) &  13.164\,(0.02) & 12.276\,(0.02) & 2.57   \\
L\,1622--6       &  05540189+0140272 & 15.862\,(0.02) &  14.464\,(0.02) & 12.635\,(0.02) & 2.45   \\
L\,1622--6 N     &  05540201+0140558 & 18.421\,(0.06) &  16.866\,(0.04) & 14.882\,(0.03) & 3.24   \\
L\,1622--8       &  05540718+0139007 & 17.424\,(0.05) &  15.190\,(0.04) & 13.584\,(0.03) & 1.88   \\
LkH$\alpha$\,336 &  05542008+0142565 & 14.229\,(0.04) &  13.188\,(0.04) & 12.278\,(0.03) & 1.70   \\
LkH$\alpha$\,336c & 05542031+0142518 & 18.160\,(0.08) &  16.707\,(0.08) & 15.416\,(0.06) & 3.67   \\ 
L\,1622--10 A    &  05542669+0152154 & 17.899\,(0.08) &  16.467\,(0.08) & 14.857\,(0.06) & 4.83   \\ 
L\,1622--10 B    &  05542696+0152172 & 18.162\,(0.08) &  16.768\,(0.06) & 15.457\,(0.05) & 6.05   \\
L\,1622--11      &  05543040+0150593 & 17.596\,(0.05) &  16.174\,(0.03) & 14.337\,(0.03) & 1.80   \\ 
LkH$\alpha$\,337 &  05543793+0129513 & 14.406\,(0.02) &  13.589\,(0.02) & 12.507\,(0.02) & 0.72   \\ 
L\,1622--15      &  05571611+0229074 & 15.650\,(0.02) &  14.974\,(0.02) & 13.950\,(0.02) & 1.66  \\ 
\hline
\end{tabular}}
\end{center}
\end{minipage}
\end{table*}

\subsection{Infrared archive data}

\subsubsection{IRAC observations}  

We measured the IRAC fluxes of the program stars 
using images from \textit{Spitzer Space Observatory} archive (Prog ID 
43: An IRAC Survey of the L\,1630 and L\,1641 (Orion) Molecular Clouds -- 
Fazio et al.). We created the IRAC mosaics from Basic Calibrated Data 
(BCD) frames in each channel using a custom IDL script which also 
converted the MJy\,sr$^{-1}$ units to DN\,s$^{-1}$ using conversion factors 
0.1088, 0.1388, 0.5952, and 0.2021 for channels 1, 2, 3, and 4, 
respectively, and performed aperture photometry on all images with 
PhotVis version 1.10, which is an IDL-GUI based photometry visualization 
tool \citep{Gutermuth04}. The radii of the source aperture and the 
inner and outer sky annuli were at 2.4~arcsec, 2.4~arcsec, and 7.2~arcsec, 
respectively. We calculated the standard IRAC magnitudes using 
$-2.5 \log$(flux$_{DN/s}$)+zp, where the zero-point term, zp, is 21.9929, 
21.2583, 19.0747, and 19.4372 for channels 1--4, respectively. The zero 
point includes the zero point of the flux to magnitude transformation 
based on the large aperture measurements of a standard star (19.66, 
18.94, 16.88, and 17.39 mag for channels 1--4); and the aperture 
corrections to account for the difference between the aperture sizes 
used for the standard star and for L\,1622 photometry ($-$0.21, $-$0.23, 
$-$0.35, and $-$0.50 mag for channels 1--4). See \citet{Reach05} for the 
detailed description of the calibration of the IRAC data.
We converted the magnitude into flux units using the equation 
$f = c\times10^{-0.4*m}$ where {\em f\/} is the flux {\em m\/} is the 
corresponding magnitude and {\em c\/} 
is the zero point in Jy (280.9, 179.7, 115.0, 64.13 for channels 1, 2, 
3, and 4 respectively). The resulting IRAC magnitudes are listed 
in Table~\ref{tab_spitzer}. L\,1622--3, L\,1622--15, and LkH$\alpha$\,334 
are located outside the field of view of the IRAC observations. 

\begin{table*}
\begin{minipage}{126mm}
\begin{center}
\caption{IRAC and MIPS magnitudes of the pre-main sequence stars of L\,1622}
\label{tab_spitzer}
{\small
\begin{tabular}{lccccc}
\hline
Star  & [3.6] & [4.5] &  [5.8] &  [8.0] & [24]    \\
\hline
LkH$\alpha$\,334 &  $\cdots$       &    $\cdots$      & $\cdots$        &   $\cdots$       & ~3.95\,(0.01)  \\
LkH$\alpha$\,335 & ~7.290\,(0.001) &  ~6.783\,(0.002) & ~6.287\,(0.001) &  5.289\,(0.001)  & 2.52\,(0.01)   \\
L\,1622--6       & ~9.316\,(0.004) &  ~9.147\,(0.003) & ~8.967\,(0.005) &  8.492\,(0.006)  & 3.88\,(0.01)  \\
L\,1622--6 N     & 10.753\,(0.003) &  10.335\,(0.003) & ~9.971\,(0.005) &  9.226\,(0.005)  & 6.78\,(0.01)  \\
L\,1622--8       & ~9.827\,(0.002) &  ~9.470\,(0.002) & ~8.934\,(0.003) &  8.069\,(0.002)  & 4.96\,(0.01)  \\
LkH$\alpha$\,336 & ~8.374\,(0.003) &  ~7.969\,(0.002) & ~7.542\,(0.002) &  6.541\,(0.004)  & 3.52\,(0.02)  \\
LkH$\alpha$\,336c& ~9.612\,(0.005) &  ~9.196\,(0.005) & ~8.632\,(0.003) &  7.644\,(0.003)  & 2.95\,(0.02)  \\
L\,1622--10 A    & ~8.838\,(0.002) &  ~8.358\,(0.002) & ~7.817\,(0.004) &  7.068\,(0.002)  & 3.11\,(0.02)  \\
L\,1622--10 B    & ~9.620\,(0.010) &  ~9.160¬,(0.010) & ~8.570\,(0.020) &  7.700\,(0.010)  & 3.00\,(0.02)  \\
L\,1622--11      & 10.618\,(0.002) &  10.289\,(0.003) & ~9.926\,(0.005) &  9.201\,(0.004)  & 6.36\,(0.01)  \\
LkH$\alpha$\,337 & ~9.679\,(0.002) &  ~8.771\,(0.003) & ~8.577\,(0.003) &  7.548\,(0.002)  & 4.33\,(0.01)   \\
IRAS~05522+0146  & 12.491\,(0.005) &  11.190\,(0.004) & 10.056\,(0.005) &  9.095\,(0.003)  & 3.69\,(0.01)  \\
\hline
\end{tabular}}
\end{center}
\end{minipage}
\end{table*}	   

\subsubsection{MIPS data} 

We also present observations of L 1622 at 24 and 70~\micron, obtained
with the Multiband Imaging Photometer for Spitzer (MIPS) as part of program
PID 43 (AOR 16214784, Nov 4 2005). The data were downloaded from the
Spitzer Science Center Archive with Leopard.  The cloud was imaged using a
single scan map with 4 legs at full array cross-scan offset, resulting
in half array gaps at 70~\micron. The scanlegs were half degree in length.
Observations were carried out using medium scan mode, resulting in a
total effective exposure time of 40~s/pixel at 24, and 20~s/pixel at 70\,\micron.

All data were processed using the MIPS instrument team Data Analysis Tool
\citep[DAT,][]{gordon05} as described by \citet{engelbracht07} and
\citet{gordon05}. Care was taken to minimize instrumental artifacts. We
used the MIPS calibration star HD 173398 as a PSF standard for the 24~\micron \
photometry, with the final PSF constructed from 72 individual AORs, kindly 
provided to us by C. Engelbracht.

Standard {\sevensize IRAF} tasks {\tt phot} and {\tt allstar} of the {\sevensize DAOPHOT}
package were used to carry out
the 24~\micron \ PSF and 70~\micron \ aperture photometry. Since the
MIPS psf is very stable and the nebulosity effects were negligible at 24~\micron, we
used a large PSF radius of 112~arcsec,
with fitting radius of 5.7~arcsec. The instrumental number flux counts
were converted to fluxes with the conversion 
1.068$\times10^{-3} ~{\rm mJy}~{\rm arcsec}^{-2}~{\rm MIPS\_UNIT}^{-1}$,  
as determined by the MIPS team \citep{engelbracht07}.
We then translated these to [24] magnitudes taking 7.17~Jy for the [24]
magnitude zero point.

The large PSF radius used ensured us that the aperture
correction was smaller than 0.06 mag at [24] (the aperture
correction for 35-arcsec aperture radius; \citealt{engelbracht07}). The correction 
factor for the PSF used has been determined
by the MIPS team during previous work (G\'asp\'ar et al. 2008, in prep.)
to be only 0.0368 magnitudes (3.4\%). We summarize our 24 photometry results 
in Table~\ref{tab_spitzer}.  L\,1622--3 and L\,1622--15 are located 
outside the field of view of the
MIPS observations. Three stars could be identified
in the 70~\micron \ images, namely LkH$\alpha$~335, LkH$\alpha$~336c, and 
L\,1622--10A. We performed aperture photometry for these stars using a 18~arcsec 
aperture radius. The sky contamination was estimated in  an annulus outside the
aperture with inner and outer radius  of  18~arcsec and 36~arcsec
respectively.  We converted the MIPS unit to Jy using a calibration
factor of 16.52 mJy/square arcsec resulting in fluxes of  529$\pm$341~mJy, 
927$\pm$404~mJy, and 504$\pm$302~mJy for LkH$\alpha$~335,
LkHalpha336c, and [UTF-8?]L\,1622--10A.respectively. No useful stellar data could
be extracted from the 160 \micron \ image.

... 

\subsection{$^{\bmath{12}}$CO and $^{\bmath{13}}$CO maps of L\,1622}
\label{Sect_CO}

Observations were carried out in the $^{12}$CO($J$ = 1--0)
emission toward the Orion~B region with the 4-m millimeter-wave
telescope, NANTEN, at the Las Campanas Observatory, from November 2001
to March 2002. The half-power beam-width (HPBW) was 2.6~arcmin at 115~GHz.
The front-end was a superconducting mixer followed by HEMT amplifiers.
Details of this mixer receiver
are given by \citet{oga90}.  The backend was an acousto-optical
spectrometer with a velocity resolution of 0.1~km\,s$\rm^{-1}$.
The telescope pointing was measured to be accurate within 20~arcsec
by observing planets.
The spectral intensities were calibrated by employing the standard
room-temperature chopper-wheel
technique \citep[][]{kut81}. The absolute intensity calibration was
made by observing Orion-KL
[R.A.(B1950) = $5^\rmn{h}32^\rmn{m}47.5^\rmn{s}$, Decl. (B1950) = $-5\degr24\arcmin21\arcsec$],
by assuming its absolute temperature, $T^{*}_{R}$  $\sim$ 65 K. We also observed
it every 2~hr to confirm the stability of the whole system.
Most of the observations were made by the frequency switching
technique with a frequency interval of 15~MHz,
while the position switching technique was also used when the
atmospheric frequency is close to the celestial $^{12}$CO frequency.

These observations were made as a part of the large scale $^{12}$CO survey of
the Orion-Eridanus region which consists of $\sim$~100,000 observing
points.
In the Orion region, $ b> -30\degr$, data were obtained
at a grid spacing 4~arcmin.  The r.m.s. noise fluctuations were 
$\sim$~0.53\,K per channel for a typical integration time of $\sim$~8\,s.

The $^{13}$CO ($J$ = 1--0) observations were made at 4-arcmin
grid spacing toward the same region as a part of large scale $^{13}$CO
survey of the Orion region
in January and February 2002.  The half-power beam-width (HPBW)
was 2.7~arcmin at 110~GHz, and the r.m.s. noise fluctuations were 
$\sim$~0.16\,K per channel.
The frequency-switching technique with a frequency interval of 15~MHz
was also used. We observed Orion-KL every 2~hr to
check the stability and establish the absolute intensity scale. We
assumed its absolute temperature  $T^{*}_{R} \sim$~10\,K.

\begin{figure}
\resizebox{\hsize}{!}{\includegraphics{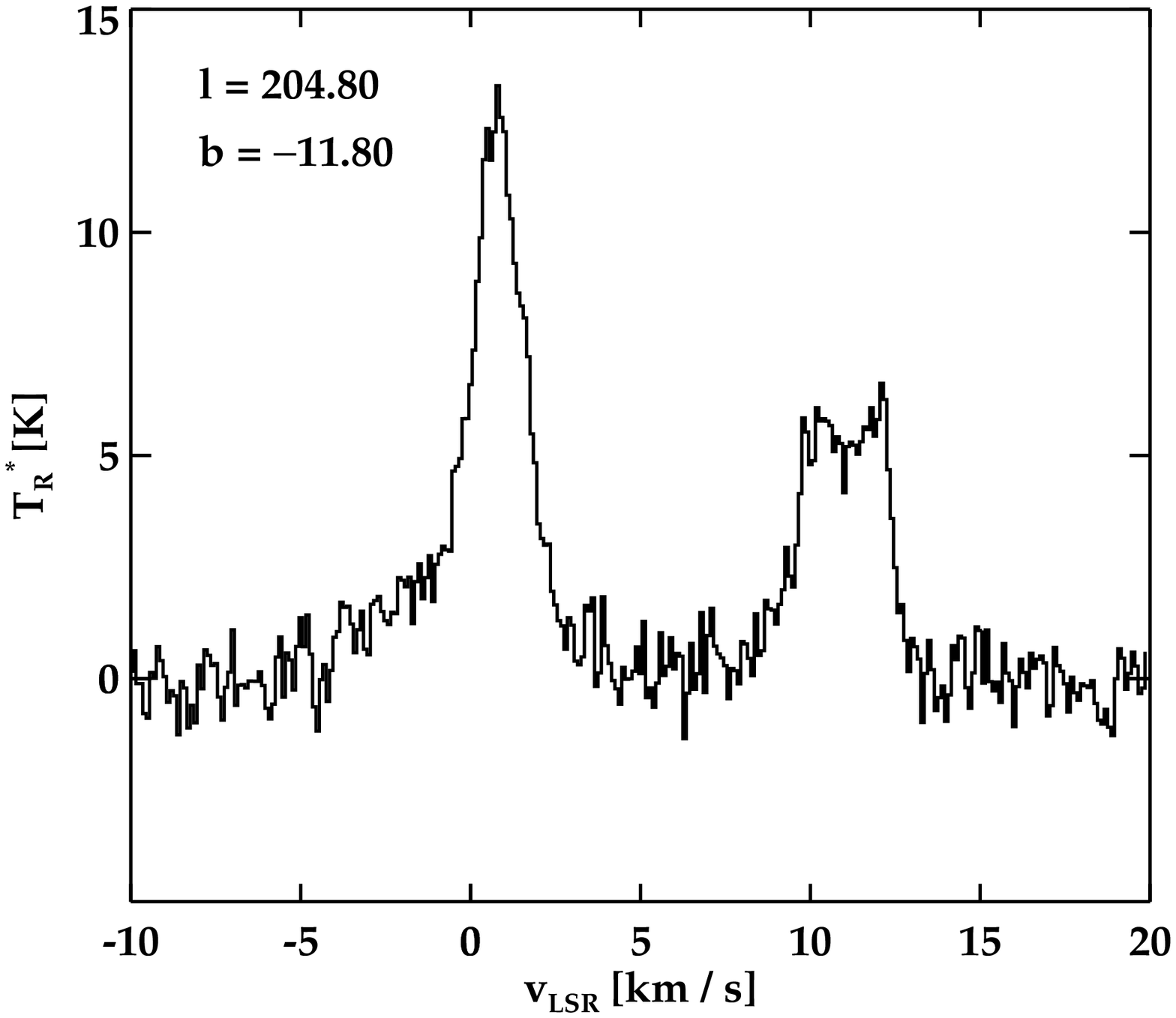}\includegraphics{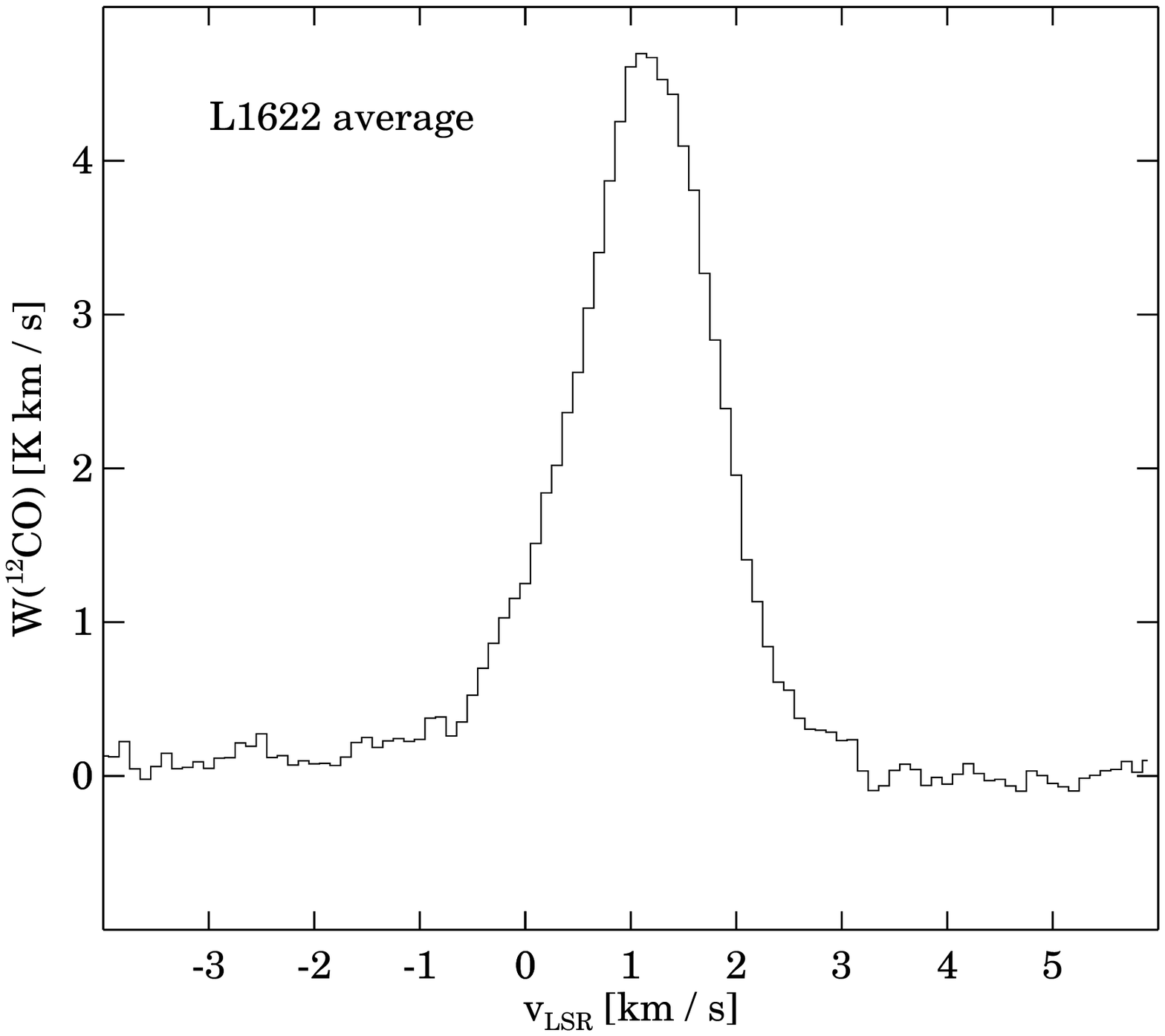}}
\caption{\textit{Left\/}: The $^{12}$CO spectrum observed at ($l,b$)=($204\fdg8,-11\fdg8$);
\textit{Right\/}: The average $^{12}$CO spectrum of L\,1622.}
\label{Fig_sp_co}
\end{figure}

The left panel of Fig.~\ref{Fig_sp_co} shows the $^{12}$CO spectrum observed 
at ({\it l,b\/})=($204\fdg8,-11\fdg8$).
The self-absorbed feature at 8--13\,km\,s$^{-1}$ corresponds to Orion~B,
whereas the peak at 1\,km\,s$^{-1}$ represents L\,1622. In the right panel 
the average of all $^{12}$CO spectra within the 3-$\sigma$ boundary of the $^{12}$CO cloud,
corresponding to L\,1622, is plotted.  The peak velocity of this
spectrum is $+1.17$\,km\,s$^{-1}$, and the line width at half maximum 
is $\Delta v = 1.66$\,km\,s$^{-1}$.

\begin{figure}
\resizebox{\hsize}{!}{\includegraphics{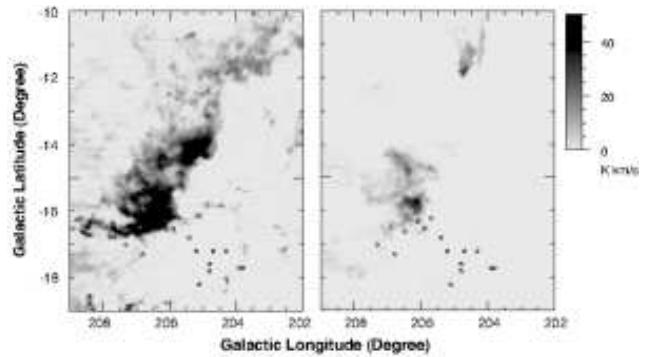}}
\caption{Large-scale distribution of the molecular gas in the region of Orion~B.
\textit{Left\/}: $^{12}$CO emission integrated over the velocity interval
 5\,km\,s$^{-1} < v_\mathrm{LSR} < 15$\,km\,s$^{-1}$  for the Orion~B main cloud. 
\textit{Right\/}: $-5$\,km\,s$^{-1} < v_\mathrm{LSR} < +5$\,km\,s$^{-1}$ map, containing
the L\,1622 molecular cloud. Star symbols indicate the O--B3 type members of Orion OB1b from
\citet{Brown94}, to show the relation between molecular clouds and Ori~OB1b.}
\label{Fig_co}
\end{figure}

The large-scale distribution of the molecular gas is displayed in Fig.~\ref{Fig_co}. 
The left panel shows the distribution of the $^{12}$CO emission integrated over the 
velocity interval of (5,15)\,km\,s$^{-1}$, and the right panel shows the same distribution
over the velocity interval ($-$5,+5.0)\,km\,s$^{-1}$.
Using the conversion formula 
$N$(H$_2$) / $W$(CO) = 1.8$\times10^{20}$\,cm$^{-2}$ / K\,km\,s$^{-1}$ \citep{Wilson05}, 
our $^{12}$CO data allow us to estimate the mass of the cloud. 
Summing up the obtained column densities within the 3-$\sigma$ contour
(1.5~K\,km\,s$^{-1}$) of the cloud, and adopting a mean molecular weight of 2.333,
the result is $M_\rmn{cloud} \approx 1100~M_{\sun}$, if we assume a distance of 400~pc.
At a distance of 120\,pc the measurements result in a mass of 100\,M$_{\sun}$.
The virial mass of the cloud, 
$M_\rmn{VIR} = 190\times R \times \Delta v^2\approx 1500~M_{\sun}$ at 400\,pc,
where $R$ is the average radius derived from the 3-$\sigma$ area of the cloud.

The mass of the $^{13}$CO emitting gas is $M_\rmn{LTE}(^{13}\rmn{CO}) = 750~M_{\sun}$, and 
 $M_\rmn{VIR}(^{13}\rmn{CO}) = 770~M_{\sun}$.  Thus the ratio M($^{13}$CO) / M($^{12}$CO) $\approx 70\%$,
a typical value for nearby dark clouds \citep{deGeus,Mizuno95,Mizuno98}.
$M_\rmn{VIR} / M_\rmn{LTE} \sim 1$  for both the $^{12}$CO and the $^{13}$CO clouds.

\begin{figure}
\resizebox{\hsize}{!}{\includegraphics{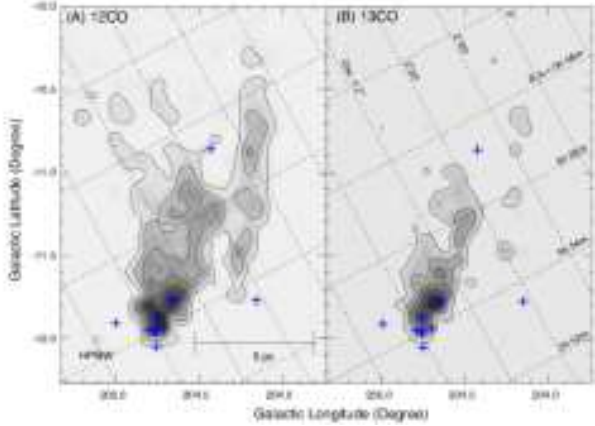}}
\caption{\textit{Left\/}: The new $^{12}$CO map of L\,1622, with the positions of PMS stars 
studied in this paper, and a grid of equatorial (J2000) coordinates  overplotted.
The lowest contour is at 1.5\,K\,km\,s$^{-1}$, and the increment is 5\,K\,km\,s$^{-1}$.
\textit{Left\/}: $^{13}$CO map of L\,1622. The lowest contour and the steps are 
1\,K\,km\,s$^{-1}$ and 2\,K\,km\,s$^{-1}$, respectively. The emission is integrated 
over the velocity interval $-5$\,km\,s$^{-1} <V_\rmn{LSR} <+5$\,km\,s$^{-1}$. in both maps.}
\label{Fig_cloud}
\end{figure}

The column density of the molecular hydrogen toward the $^{13}$CO peak at
$(l,b) = (204.8, -11.8)$ is N(H$_2$) = $9.3\times10^{21}$\,cm$^{-2}$. The typical column 
densities of the isolated small $^{13}$CO clouds around the main cloud 
are about N(H$_2$) = $1.9\times10^{21}$\,cm$^{-2}$.
Figure~\ref{Fig_cloud} shows the $^{12}$CO and $^{13}$CO maps of L\,1622, with
the positions of the associated PMS stars overplotted.

\section{Discussion}
\label{Sect_3}

\subsection{Description of the region}
\label{Sect_3.1}

L\,1622 harbours two \textit{IRAS} point sources whose flux ratios are different from 
those of ordinary stars. \textit{IRAS}~05517+0151 coincides with a visible star. 
Both the flux ratio $F_{12}/F_{25}$  and the near-infrared
properties of its  associated 2MASS source, 2MASS\,J\,05542277+0152039, 
are indicative of an evolved star. The optically invisible \textit{IRAS}~05522+0146 is probably 
an embedded young stellar object. Within the uncertainty ellipse it coincides in 
position with 2MASS~J\,05545377+0147098.
This source was not detected at 12\,$\mu$m by \textit{IRAS}, but the high upper limit of the
12\,$\mu$m flux, 0.34\,Jy, is compatible with the assumption that the \textit{IRAS} 
and 2MASS fluxes originate from the same protostellar source. 
No infrared source can be found at the position of  
HH~122~VLA\,1 \citep{Rodriguez}. 

\begin{figure}
\resizebox{\hsize}{!}{\includegraphics{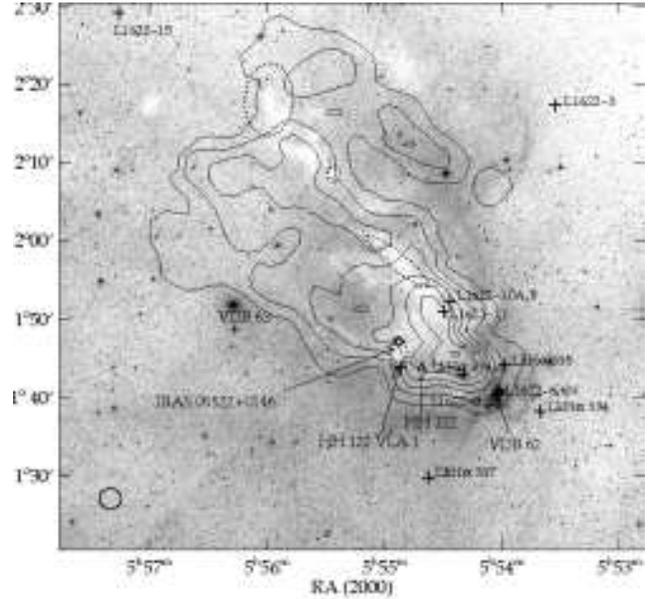}}
\caption{Distribution of the $^{12}$CO emission and the young stellar objects,
overplotted on the DSS2 red image of the L\,1622 region. The value of the lowest  
contour plotted is 1.5\,K\,km\,s$^{-1}$, and the
increment is 5\,K\,km\,s$^{-1}$. Crosses indicate 
the pre-main sequence stars, and dotted ellipses show the
optically selected cores \citep{LM99}. Position of HH\,122, HH\,122~VLA\,1, 
and \textit{IRAS}~05522+0146, as well as the reflection nebulae VDB~62 and VDB~63  
are also indicated. The beam size of the $^{12}$CO observations is shown by the circle
at the lower left corner.}
\label{Fig_map}
\end{figure}

Figure~\ref{Fig_map} shows the surface distribution of the $^{12}$CO emission and 
star formation signposts, overplotted on the DSS\,2 red image of L\,1622. 
The value of the lowest $^{12}$CO contour plotted is 1.5\,K\,km\,s$^{-1}$, and the
increment is 5\,K\,km\,s$^{-1}$. 
Crosses indicate the pre-main sequence stars, and dotted ellipses show the
optically selected cores \citep{LM99}. Positions of HH\,122, HH\,122~VLA\,1, 
and \textit{IRAS}~05522+0146, as well as the reflection nebulae VDB~62 and VDB~63  
are also indicated. 

The $^{12}$CO distribution shows a cometary shaped cloud with a bright rimmed head
region on the southwestern side. This shape suggests that star formation might
have been triggered by a shock arriving from the southwest, where the luminous stars 
of Ori~OB1 can be found.  Four pre-main sequence stars, 
namely L\,1622--3, LkH$\alpha$\,334, LkH$\alpha$\,337, and L\,1622--15 are
projected outside the observed molecular cloud, thus their relation to L\,1622
is uncertain. In particular, L\,1622--3 and L\,1622--15, located well outside the 
lowest contour of the tail of the cloud, may be unrelated 
to L\,1622. LkH$\alpha$\,334 and LkH$\alpha$\,337 are projected near the 
head of the cloud, on the southwestern side, and thus might have been born in L\,1622 at
an earlier epoch when the cloud was larger. 
On the other hand, these isolated stars might have been formed in small clouds 
around the main  large cloud, as discussed in \citet{Mizuno98} for the Chamaeleon  
region.  Ten PMS stars are projected within the 
head of the cloud. Four of them,  HD\,288313, L\,1622--6, L\,1622--8 and 
L\,1622--6N form a compact group near the bright rim. Three 
stars, L\,1622--10\,A, L\,1622--10\,B, and  L\,1622--11, are projected 
on a dark spot within the cloud, associated with the $^{13}$CO peak. Both 
candidate embedded protostars,
HH~122~VLA\,1 and \textit{IRAS}~05522+0146, are associated with the core L\,1622\,B. 

\subsection{Positions of the pre-main sequence stars in the HRD}
\label{Sect_3.4}

\begin{figure*}
\centerline{\includegraphics[width=7cm]{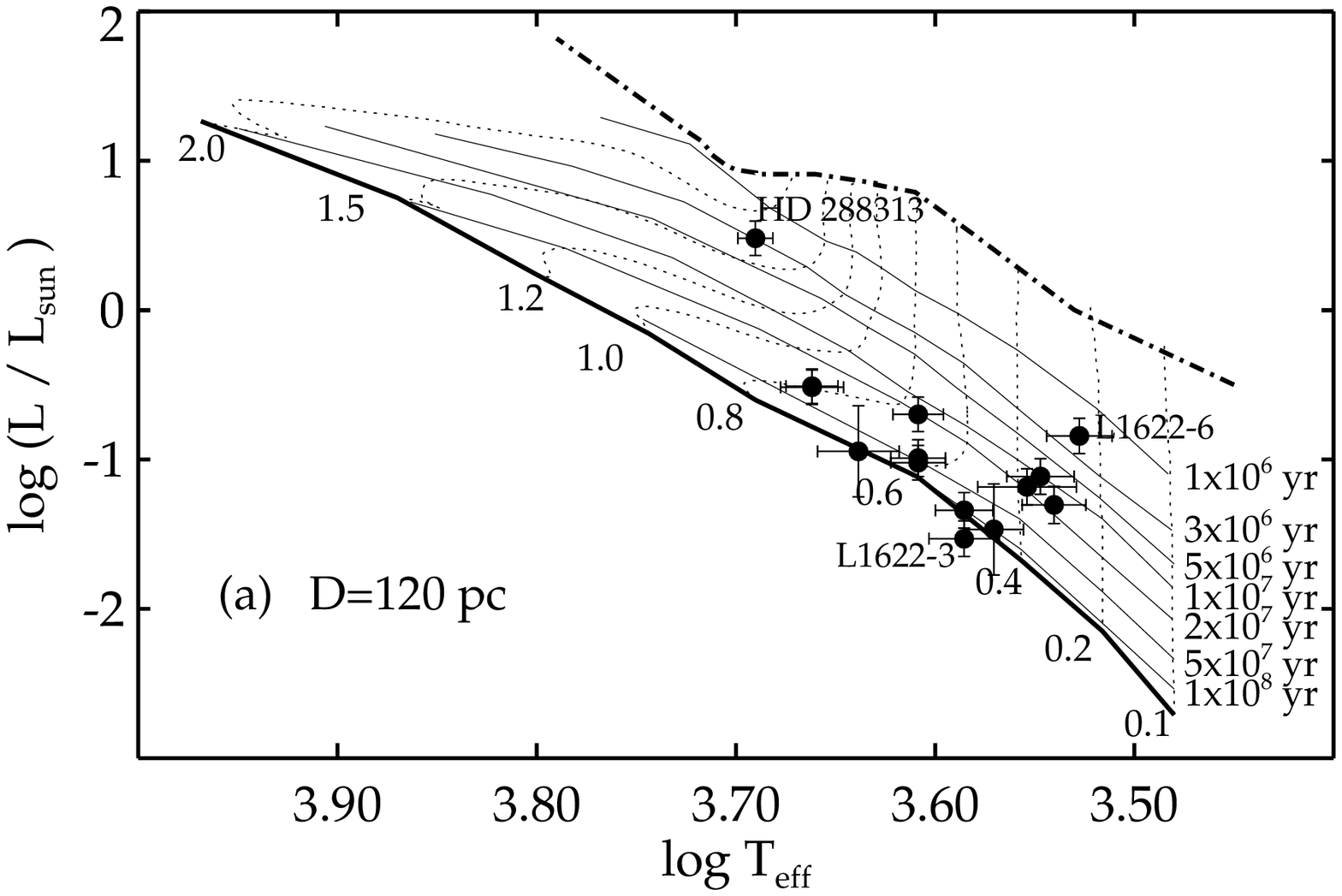}\hskip3mm\includegraphics[width=7cm]{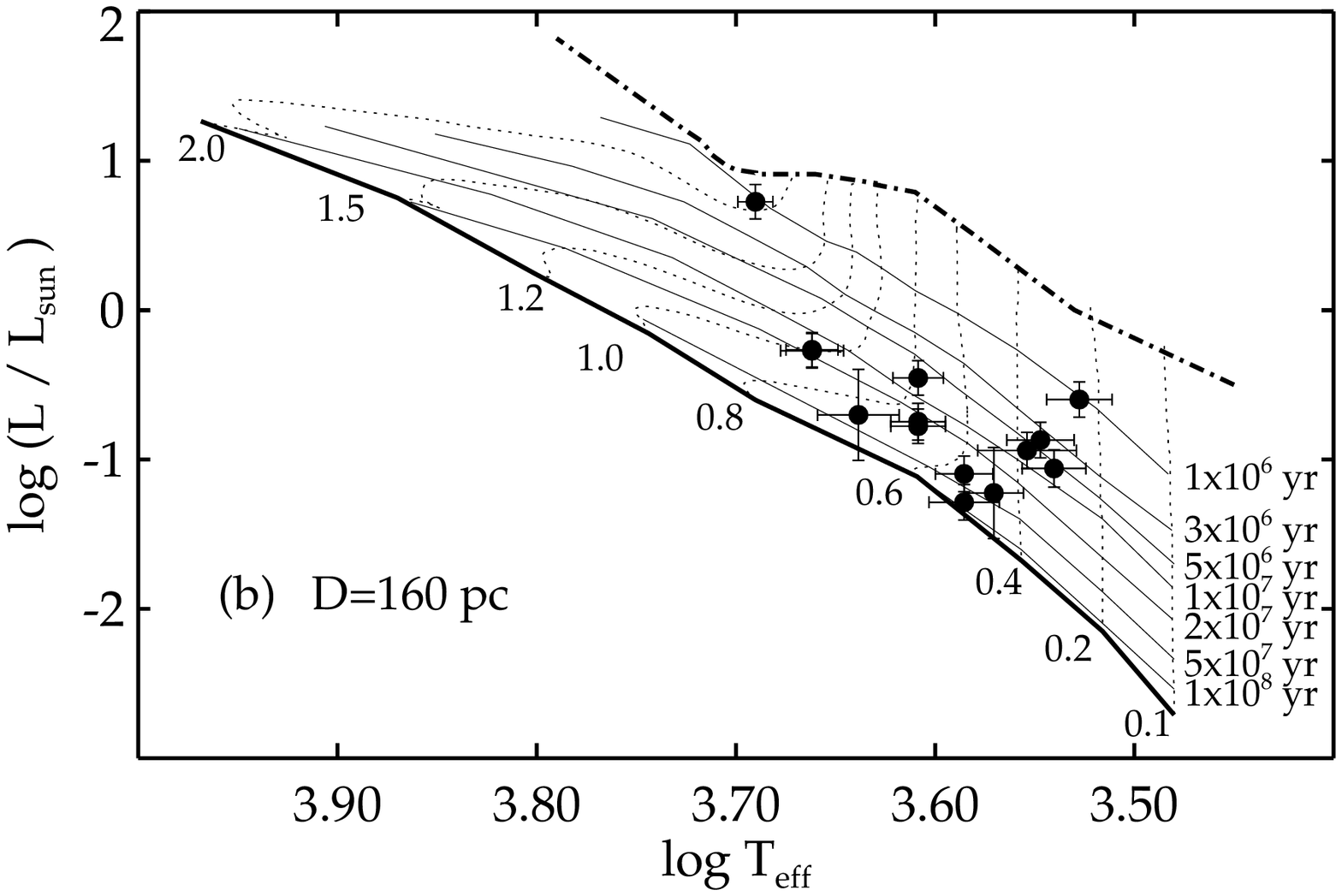}}
\centerline{\includegraphics[width=7cm]{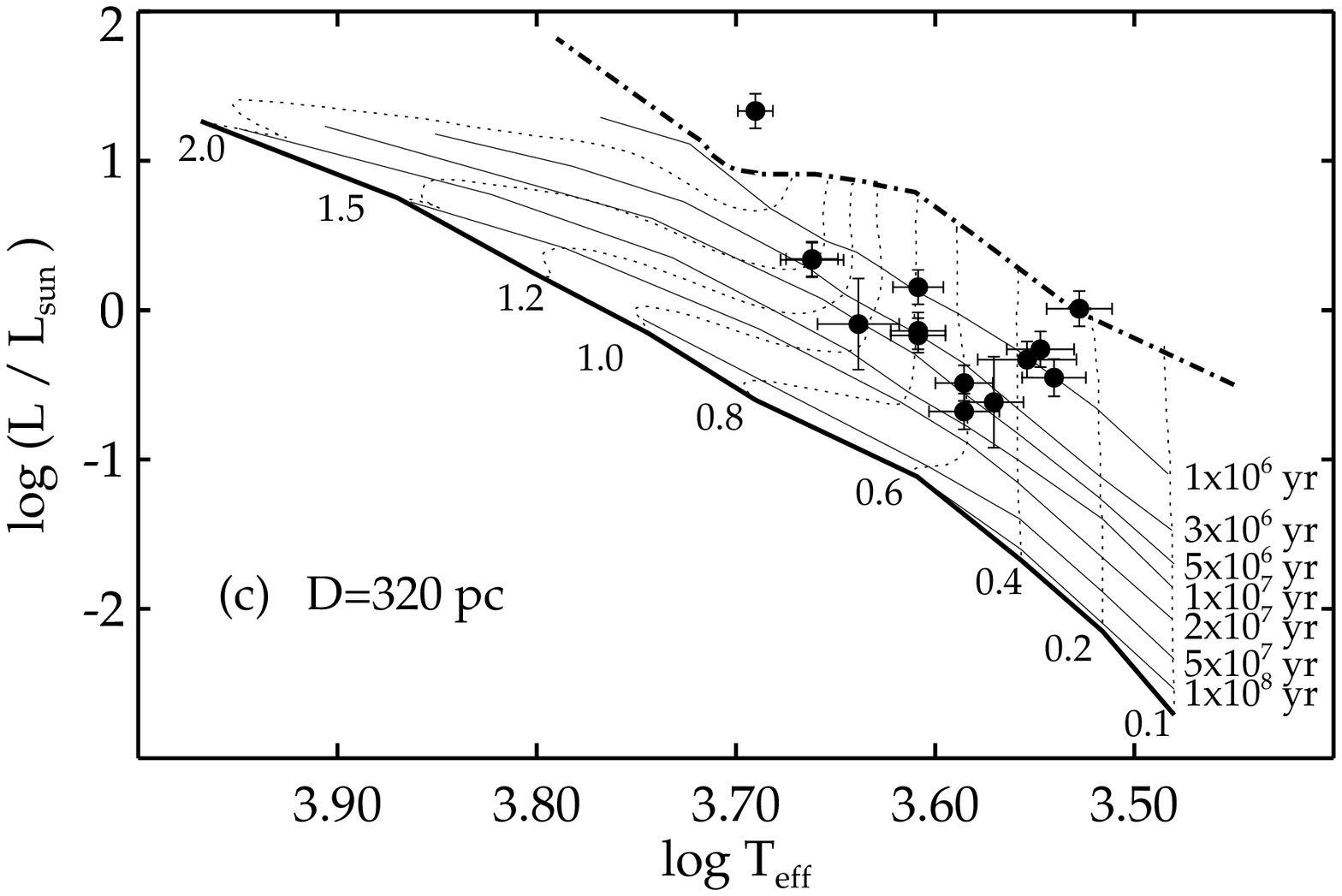}\hskip3mm\includegraphics[width=7cm]{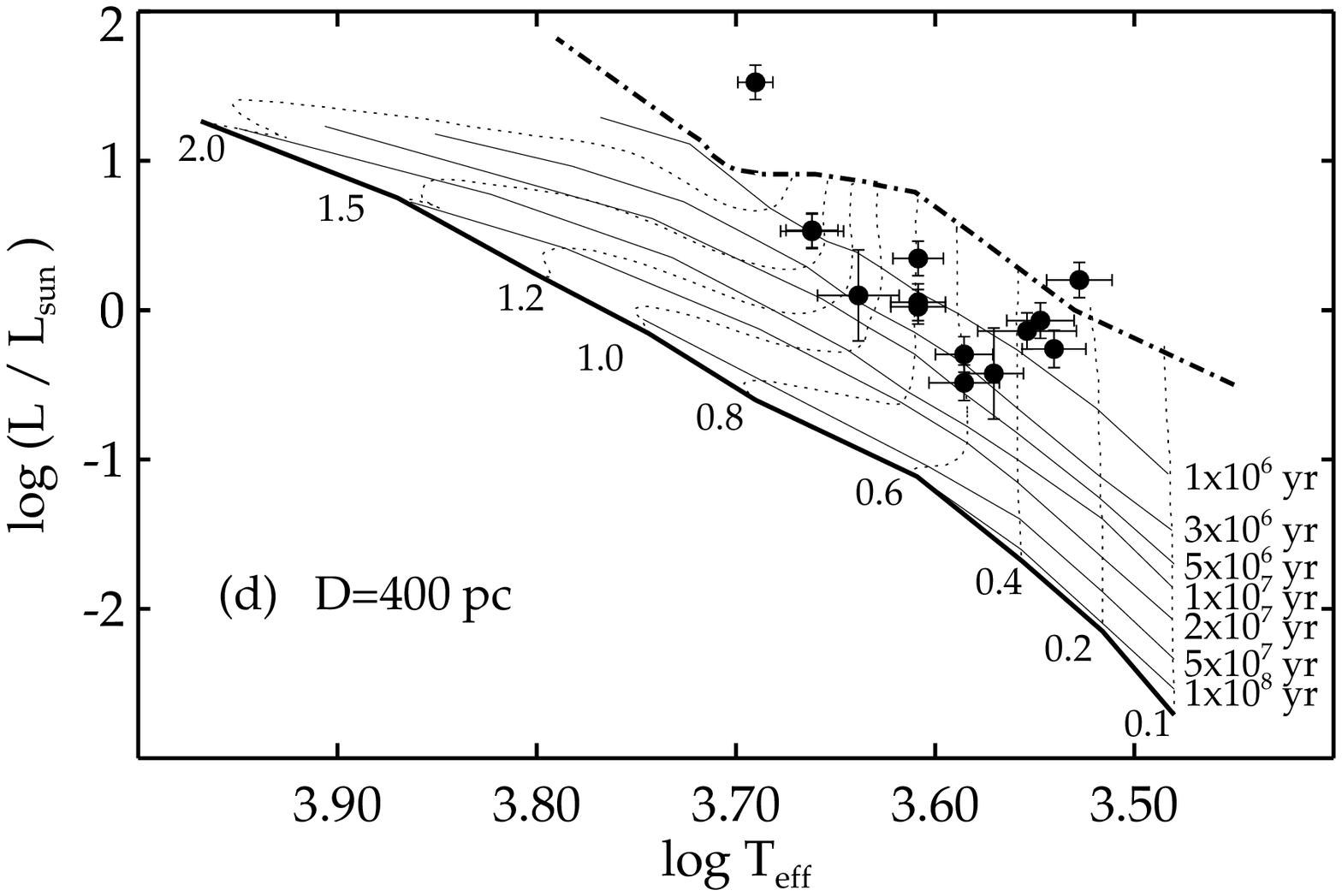}}
\caption{Positions of the observed stars in the HRD assuming that the distance
of the cloud is (a) 120\,pc, (b) 160~pc, (c) 320~pc, (d) 400~pc.
Thin solid lines indicate the isochrones as labelled,
and dotted lines show the evolutionary tracks for the masses (M$_{\sun}$) indicated at the 
lower end of the tracks according to \citet{PS99} model. The dash-dotted line
corresponds to the birthline and thick solid line indicates the
zero age main sequence.}
\vskip -0.4cm
\label{Fig_hrd}
\end{figure*}

The Hertzsprung--Russell diagrams (HRD) of the pre-main sequence stars of L\,1622, 
for four different distances, are displayed in Fig.~\ref{Fig_hrd}. Error bars of 
$\log\,T_\rmn{eff}$ are derived from the accuracy 
of the spectral classification. The error of the 
luminosity comes from the quadratic sum of $\delta I_{C}$, 
$\delta A_{I_{C}}$,  and $\delta BC_{I_{C}}$.  
Evolutionary tracks and isochrones, as well as the position of the 
birthline and zero-age main-sequence \citep{PS99} are also shown.

Most of the programme stars are distributed over a fairly narrow age range.
Exceptions are  HD\,288313 and  L1622--6, which are apparently far more  luminous 
than the other stars. Since HD\,288313 is a multiple system \citep{Reipurth08}, 
the observed luminosity has to be distributed on its components. L1622--6
is either a foreground object, or, more probably, a multiple system as well.
The distribution of the stars with respect the isochones shows that the extreme 
emission line star L\,1622--3 is apparently the oldest member
of the sample, in contrast with the observations that PMS activities decrease
with stellar age. It is unlikely that the relatively low luminosity of this star is due to
its larger age. L\,1622--3 may either be a higher-mass counterpart of the 
underluminous, highly accreting very low mass stars LS--RCrA~1, Par-Lup3--4 \citep{Comeron} 
and KPNO~Tau~12 \citep{Luhman03}, or may be somewhat more distant than L\,1622. 
Its position on the sky with respect to the cloud suggests that L\,1622--3 might 
have been born in another small cloud, projected near L\,1622.  Radial velocity 
measurements may clarify its  connection with L\,1622. 

At a distance of 120~pc \citep[panel (a),][]{Wilson05}, most stars are 
close to the ZAMS, while L1622--3 and
L1622--15 are on the ZAMS. The average age of the stars, suggested by 
this distance, is not compatible with their CTTS nature and triggering of star 
formation in L\,1622 from the older subgroups of Ori~OB1.
The same is true for 160~pc, shown in panel (b): the mean age of the stars
would be about 20 million years at this distance. At 320~pc (panel (c)), suggested 
by the photometric properties of VDB\,63, the mean age of the pre-main sequence
stars is about 3~million years, whereas, if L\,1622 is situated at 400~pc from us, 
the age of its YSO population is around 1~million years (panel (d) of  Fig.~\ref{Fig_hrd}).
 We propose to accept 400~pc for the distance of L\,1622, and thus the age of $\la 1$~million year 
for the average age of its PMS star population. This age is compatible with that suggested
by the spectroscopic properties of the stars, i.e. the high proportion of accretors (85\%)
among the H$\alpha$ emission stars \citep[cf.][]{Jaya06}, as well the high proportion of
the protostars among the YSO population \citep[10/28,][]{Reipurth_hb}.
 
\subsection{Spectral energy distributions of the target stars}
\label{Sect_sed}

\begin{figure*}
\centering{\includegraphics[width=14cm]{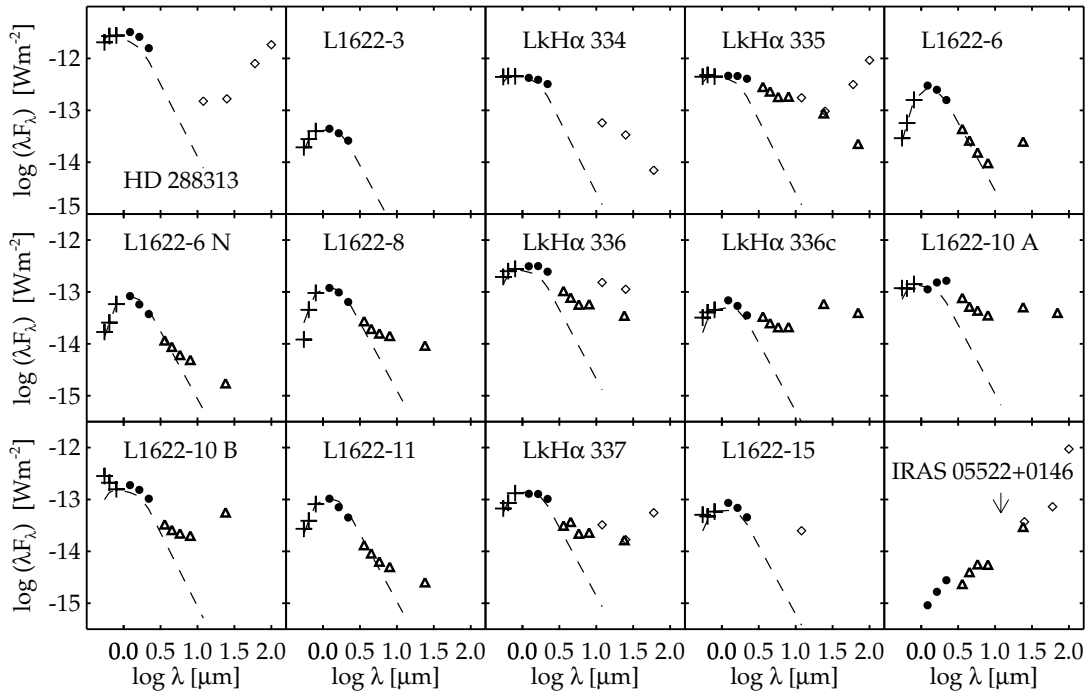}}
\caption{Spectral energy distributions of the PMS stars and \textit{IRAS}~05522+0146, based on
our optical photometry (plusses),  2MASS (dots), \textit{Spitzer} IRAC and MIPS (triangles), 
as well as  \textit{IRAS} (thin diamonds) data. The arrow in the last panel
indicates the upper limit of the 12-micron \textit{IRAS} flux of the source.}
\label{Fig_sed}
\end{figure*}

The shape of the SED, reflecting the structure of the circumstellar matter of the YSOs,
is a useful indicator of evolutionary state \citep[e.g][]{Lada84,ALS87}. 
The optical, near- and mid-infrared parts of the spectral energy distributions of the programme
stars were constructed using the $VR_{C}I_{C}JHK_s$ magnitudes and IRAC and MIPS fluxes, 
each corrected for the interstellar extinction. The fluxes corresponding to 
zero magnitude were obtained from \citet{Glass} for the $V R_{C} I_{C}$ bands, 
and from the 2MASS All Sky Data release web 
document\footnote{$\rmn{http://www.ipac.caltech.edu/2mass/releases/allsky/doc/}$
 $\rmn{sec6\_4a.html}$}
for the $J H K_s$ bands. \textit{IRAS} fluxes are also plotted when available  \citep{Weaver}.
The resulting  SEDs are shown in Fig.~\ref{Fig_sed}.
Photospheric SEDs have been drawn by dashed lines. 
Their values were determined from the dereddened $I_{C}$ magnitudes, and from
the colour indices corresponding to the spectral types. The right bottom panel of 
Fig.~\ref{Fig_sed} shows the fluxes of \textit{IRAS}~05522+0146, with the assumption that it is
identical with 2MASS\,J\,05545377+0147098. In this case the fluxes are not corrected for the
interstellar extinction. The tip of the downward arrow indicates the upper
limit of its 12-micron flux. 

The SEDs exhibit near-infrared ($H$, $K_s$) colour excess of LkH$\alpha$\,335, 
LkH$\alpha$\,336, and  L\,1622--10~A, indicative of disc inner edges at the dust sublimation
radius, characteristic of actively accreting stars, probably not older than 1~million years. 
The signature of the disc  appears at longer wavelengths in the SEDs of L\,1622--6,
 L\,1622--6N, L\,1622--8, and  L\,1622--11, suggesting more evolved discs around these 
stars \citep[e.g][]{Furlan06}. 
We note that these latter stars, except L\,1622--11 are members of the small compact 
group near the bright rim, whereas most of the stars with NIR excess are projected deeper 
inside the molecular cloud, suggesting star formation propagating into the cloud from the 
direction of the bright rim. We also note that the signature of the disc in the 
SED of L\,1622--6N confirms the PMS nature of this star. 

\begin{table*}
\begin{center}
\caption{Parameters of the programme stars, derived from SED model fitting and from our optical data, 
 assuming a distance of 400~pc, as well as
the main properties of the circumstellar matter, according to the best fit models of \citet{Robitaille}.
Stars without \textit{Spitzer} data are not listed.}
\label{Tab_sed}
{\small
\begin{tabular}{lc@{\hskip2mm}cc@{\hskip2mm}cc@{\hskip2mm}r@{\hskip2mm}c@{\hskip2mm}c@{\hskip2mm}c@{\hskip2mm}c@{\hskip2mm}c@{\hskip2mm}c@{\hskip2mm}c@{\hskip2mm}c}
\hline
Star  & \multicolumn{2}{c}{$T_\rmn{eff}$} & \multicolumn{2}{c}{$M_{*}$} & \multicolumn{2}{c}{Age} & $\dot{M}_\rmn{env}$ &  $R_\rmn{env}$ &
$M_\rmn{disc}$ & $\dot{M}_{d}$ & \multicolumn{2}{c}{$R_{d}^\rmn{min}$} &  $R_{d}^\rmn{max}$ & $i$ \\[3pt]
     & \multicolumn{2}{c}{(K)} & \multicolumn{2}{c}{(M$_{\sun}$)}  & \multicolumn{2}{c}{(Myr)} & (M$_{\sun}$ yr$^{-1}$) & (AU) &
     (M$_{\sun}$) & (M$_{\sun}$ yr$^{-1}$) & ($R_\rmn{sub}$) & (AU) & (AU) & ($\deg$) \\
     & SED & Sp. & SED & HRD &  SED & HRD &  \\
\hline
LkH$\alpha$\,335 & 4260 & 4590 & 1.46 & 1.60 & 0.13 & 1.0     & 1.4E$-$4 & 1.6E+4 & 8.6E$-$4 & 2.0E$-$8  & ~1.0 & 0.30 & ~~58 & 49 \\
L\,1622--6       & 3440 & 3370 & 0.33 & 0.25 & 0.25 &  0.0    & 6.5E$-$7 & 3.2E+3 & 3.3E$-$4 & 6.1E$-$11 & 48.0 & 3.14 & 1210 &  18 \\
L\,1622--6 N     & 3374 & 3470 & 0.30 & 0.30 & 1.7 & $\la1.0$ & $\cdots$ &$\cdots$  & 6.9E$-$5 & 9.0E$-$11 & ~4.4 & 0.15 & ~124 &  70 \\
L\,1622--8       & 3405 & 3525 & 0.31 & 0.35 & 1.5 & $\la1.0$ & $\cdots$ & $\cdots$ & 8.8E$-$3 & 8.6E$-$8  & ~1.0 & 0.04 & ~103 &  32 \\
LkH$\alpha$\,336 & 4120 & 4060 & 0.80 & 0.80 & 1.0 & $<1.0$   & 7.2E$-$9 & 1.9E+3 & 3.2E$-$2 & 6.8E$-$7  & ~1.4 & 0.80 & ~100 &  76 \\
LkH$\alpha$\,336c& 3980 & 3720 & 0.70 & 0.50 & 0.2 & 2.5      & 9.4E$-$7 & 4.0E+3 & 4.1E$-$3 & 2.4E$-$7  & ~1.0 & 0.16 & ~~56 &  81 \\
L\,1622--10 A    & 4010 & 4060 & 0.70 & 0.80 & 0.4 & 1.5      & 4.0E$-$7 & 1.2E+3 & 2.8E$-$4 & 5.7E$-$10 & ~1.0 & 0.14 & ~145 & 81 \\
L\,1622--10 B    & 4200 & 4350 & 1.01 & 0.90 & 0.3 & 3.0      & 9.3E$-$7 & 1.6E+3 & 6.0E$-$5 & 2.0E$-$11 & ~1.0 & 0.17 & 984 & 87 \\
L\,1622--11      & 3540 & 3580 & 0.37 & 0.35 & 1.9 & $\la1.0$ & $\cdots$ & $\cdots$ & 8.2E$-$5 & 1.6E$-$11 & ~2.9 & 0.11 & ~100 &  32 \\
LkH$\alpha$\,337 & 4030 & 4060 & 0.75 & 0.80 & 4.0 & 1.5      & $\cdots$ & $\cdots$ & 9.1E$-$3 & 3.4E$-$8  & ~4.9 & 0.30 & ~198 &  69 \\
\hline
\end{tabular}}
\end{center}
\smallskip
\flushleft{
{\small Column 1: Name of the star; Cols. 2--3: $T_\rmn{eff}$ derived from the SED model fit and spectral classification,
respectively; Cols. 4--5: Mass of the star in M$_{\sun}$, derived from the SED model fit and HRD, respectively; Cols. 6--7: age of the star,
derived from the SED model and HRD, respectively; Col. 8: accretion rate from the envelope in M$_{\sun}$ yr$^{-1}$;
Col. 9: outer radius of the envelope in AU; Col.~10: disc mass in M$_{\sun}$; Col. 11: disc accretion rate in M$_{\sun}$ yr$^{-1}$; 
Col. 12: inner radius of the disc as a multiple of the dust sublimation radius $R_{\rmn{sub}}$; Col.~13: same in AU; 
Col.~14: outer radius of the disc in AU; Col. 15: inclination (deg).}}
\end{table*}	   

In order to get a deeper insight into the circumstellar structures of the program stars,
we fitted the observed SEDs with YSO model SEDs using the large grid of radiation 
transfer models described 
by \citet{Robitaille}\footnote{Available online at http://www.astro.wisc.edu/protostars.}. 
Since  several combinations of \textit{central star--disc--infalling envelope} 
can result in good fit with the observed SED, we selected those model SEDs from the best fits that
reproduced the stellar parameter values ($A_V$, $T_\rmn{eff}$, $M_{*}$) 
determined from  our optical observations.  Table~\ref{Tab_sed} shows the 
main stellar and circumstellar parameters resulted from the model 
fitting. The  stellar parameters obtained from the optical observations are also shown 
for comparison. We have not fitted the SEDs of stars without \textit{Spitzer}
data. 

It can be seen that the best models for L\,1622--6N, L\,1622--8, L\,1622--11, and 
LkH$\alpha$\,337 contain only a star and an accretion disc, whereas 
the SEDs of  LkH$\alpha$\,335, L\,1622--6, L\,1622--10A, 
LkH$\alpha$\,336, and LkH$\alpha$\,336c could not be fitted without including an envelope.
The shape of the SED of L\,1622--6 ranks this object 
into the class of transitional discs, having a wide inner hole. Envelopes usually disappear
before development of the inner disc hole, therefore we do not expect an
envelope around objects whose SED shapes are similar to that of  L\,1622--6. 
The large inner disc hole may be a consequence of the 
binarity of this star, also suggested by its high luminosity in the HRD. 
The inner regions of the accretion disc of this object might have been evacuated  
by a close companion.  

In the case of  LkH$\alpha$\,336c and  L\,1622--10A the fits suggest 
discs seen nearly edge on. Contribution of scattered light is significant in the 
short-wavelength part of the SEDs of these stars, as we have seen in Sect.~\ref{Sect_2.3}. 
The SED of \textit{IRAS}~05522+0146  shows this object to be a Class~I source, with 
signs of strong water ice and silicate absorptions around 3 and 10~\micron, respectively.

\subsection{A possible scenario of star formation in L\,1622}
\label{Sect_scen}

The Hertzsprung--Russell diagrams presented in Fig.~\ref{Fig_hrd} suggest that
L\,1622 is not associated with the near side of the Orion--Eridanus Bubble. The
extinction layers identified at 120 and 160~pc, that represent the interaction region
of the Local Bubble and Ori--Eri Bubble, are foreground to this cloud. 
The derived parameters of young stars are compatible with both the distance 
of Ori~OB1a (330~pc) and Ori~OB1b (440~pc). The cloud is projected on Orion~B, but is 
diverging from this giant cloud toward us with a radial velocity of 9~km\,s$^{-1}$. 
Thus the true distance of the cloud from the Sun is probably 
slightly smaller than 400~pc.
This geometry suggests that the source of the high pressure which initiated star 
formation in L\,1622 may be somewhat more distant than the cloud itself. 

\begin{figure}
\resizebox{\hsize}{!}{\includegraphics{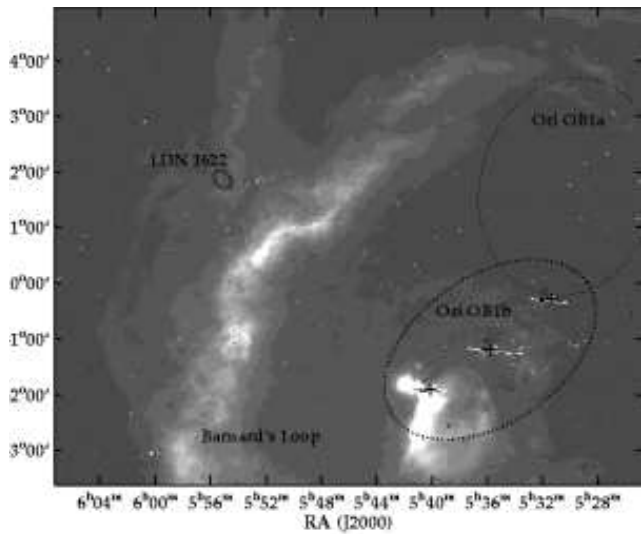}}
\caption{Continuum subtracted H$\alpha$ (\textit{SHASSA}) image of a $10\fdg6\times8\fdg6$
field in the Orion region demonstrating the position of L\,1622 (solid ellipse) 
relative to the association  subgroups Ori~OB1a and Ori~OB1b (dotted ellipses). 
Crosses indicate the Belt's brightest stars.}
\label{Fig_hamap}
\end{figure}

The bright, ionized head of the cloud points toward Orion's Belt (Fig.~\ref{Fig_hamap}),
suggesting that stellar winds and ionization front from the supergiant members of 
the association subgroup Ori~OB1b, whose age is about 6--7~million years \citep{Blaauw91,Bally08},
might have provided the shock which initiated star formation in L\,1622. The separation 
of $\sim4.8\deg$ between L\,1622 and the easternmost Belt star $\zeta$~Ori corresponds to 
some 34~pc at a distance
of 400~pc. Assuming an age difference of $\sim$\,6~million years the average
expansion velocity of the \mbox{H\,{\sc ii}} zone or stellar wind bubble should
have been about 6.8~km\,s$^{-1}$,  a reasonable value for such effects 
\citep[e.g][]{EL77,Elmegreen}.  Similar interaction between L\,1622 and Ori~OB1a, assuming a
distance of $\sim$\,320~pc, is less likely, considering the orientation of the bright rim 
of L\,1622  and the projected position of Ori~OB1a.

\section{Conclusions}
\label{Sect_concl}

We determined effective temperatures and luminosities of 14 pre-main sequence stars 
and candidates in the region of the dark cloud L\,1622 and found that these 
quantities are incompatible with the assumption that L\,1622 lies at the
near side of the Orion--Eridanus Bubble, but suggest that this cloud is located
at the distance of Orion~B. The derived ages of the stars at this distance 
are around 1 million~years. The shapes of the SEDs of the target stars support this
age estimate. Star formation in L\,1622 was probably triggered
by the luminous stars of the association subgroup Ori~OB1b (Orion's Belt). 

We present new $^{12}$CO and $^{13}$CO maps of L\,1622. Accepting a distance of 400~pc
we derived a mass of 1100\,M$_{\sun}$ for L\,1622 from the $^{12}$CO observations. The total 
mass of the YSOs identified so far, assuming a mean mass of 0.5\,M$_{\sun}$ for the
\textit{Spitzer} sources not included in our spectroscopic survey, is $\approx$20\,M$_{\sun}$, 
suggesting a star formation efficiency $SFE \approx 1.8\%$.

Several of our programme stars may be interesting targets for more detailed studies.
The derived luminosity of the CTTS L\,1622--6 suggests that this object
is a binary or multiple system. The large inner hole in the accretion disc of
this star, suggested by the models fitted to its SED, supports this hypothesis.
The SEDs of L\,1622--10A, L\,1622--10B, and LkH$\alpha$\,336c suggest that the discs of these 
stars are seen nearly edge-on. L\,1622--3  exhibits an extremely rich emission spectrum. 
This star might have been born in a small molecular clump  outside the main cloud.
\textit{IRAS}~05522+0146 is a Class~I source embedded in the dense core L\,1622\,B.

\section*{Acknowledgements}

Our results are partly based on observations obtained at the Centro 
Astron\'omico Hispano Alem\'an (CAHA) at Calar Alto, operated jointly by the 
Max-Planck-Institut f\"ur Astronomie and the Instituto de 
Astrof\'{\i}sica de Andaluc\'{\i}a (CSIC). Our observations were partly supported 
by the OPTICON project. OPTICON has received research funding from the European 
Community's Sixth Framework Programme under contract number RII3-CT-001566.
We thank Calar Alto Observatory for allocation 
of director's discretionary time to this programme. This publication makes use of 
data products from the Two Micron All Sky Survey, 
which is a joint project of the University of Massachusetts and the Infrared 
Processing and Analysis Center/California Institute of Technology, funded by 
the National Aeronautics and Space Administration and the National Science Foundation.
This work makes use of observations made with the \textit{Spitzer Space Telescope},
which is operated by the Jet Propulsion Laboratory, California Institute of Technology
under a contract with NASA. 
We thank John Bally for helpful discussions, and Bo Reipurth for communicating to us 
unpublished information. We also thank the referee, Matilde Fernandez, for her very 
constructive report which helped to improve this paper.
Financial support from the Hungarian OTKA grant T49082 is acknowledged.

\end{document}